\documentclass[11pt]{article}
\usepackage{graphicx}
\usepackage{cite}

\newcommand{ \be}{\begin{equation}}
\newcommand{ \ee}{\end{equation}}
\newcommand{ \bea}{\begin{eqnarray}}
\newcommand{ \eea}{\end{eqnarray}}
\newcommand{ \mysmall}[1]{\scriptscriptstyle #1} 

\newcommand{ \amu}{a_{\mu}}
\newcommand{ \mw}{M_{\mysmall{W}}}
\newcommand{ \mz}{M_{\mysmall{Z}}}
\newcommand{ \mh}{M_{\mysmall{H}}}

\newcommand{ \eq}[1]{eq.~(\ref{eq:#1})}
\newcommand{ \gev}  {\mbox{ GeV}}
\newcommand{ \bm}   {\boldmath}
\newcommand{ \ubm}  {\unboldmath}

\begin{document}

\begin{titlepage}
\newcommand\hepnumber{hep-ph/0411168}
\begin{flushright}
        hep-ph/0411168 \\
	Updated February 2005
\end{flushright}
\renewcommand{\thefootnote}{\fnsymbol{footnote}}

\begin{center}
\vspace{1.2cm}
{\LARGE \bf The Standard Model Prediction of the Muon\\[3mm]
Anomalous Magnetic Moment}

\vspace{1.5cm} {\Large\bf M.~Passera\footnotetext{e-mail address:
passera@pd.infn.it}}

\vspace{1cm}
{\it    Dipartimento di Fisica ``G.~Galilei'', Universit\`{a} 
        di Padova and \\ INFN, Sezione di Padova, I-35131, Padova, Italy 
	\\[3mm]
        Departament de F\'\i sica Te\`orica and IFIC Centro Mixto,
        Universitat de Val\`encia--CSIC, E-46100, Burjassot, Val\`encia, 
	Spain
}
\vspace{2.5cm}

{\large\bf Abstract} 
\end{center} 

\vspace{3mm} 
\noindent This article reviews and updates the Standard Model prediction of
the muon $g$$-$$2$. {\small QED}, electroweak and hadronic contributions are
presented, and open questions discussed.  The theoretical prediction
deviates from the present experimental value by 2--3 standard deviations, if
$e^+e^-$ annihilation data are used to evaluate the leading hadronic term.

\end{titlepage}
\setcounter{page}{2}
\tableofcontents\newpage

\section{Introduction}

The evaluation of the Standard Model ({\small SM}) prediction for the
anomalous magnetic moment of the muon $\amu \equiv (g_{\mu}-2)/2$ has
occupied many physicists for over fifty years. Schwinger's 1948
calculation~\cite{Sch48} of its leading contribution, equal to the one of
the electron, was one of the very first results of {\small QED}, and its
agreement with the experimental value of the anomalous magnetic moment of
the electron, $a_e$, provided one of the early confirmations of this theory.

While $a_e$ is rather insensitive to strong and weak interactions, hence
providing a stringent test of {\small QED} and leading to the most precise
determination to date of the fine-structure constant $\alpha$, $\amu$ allows
to test the entire {\small SM}, as each of its sectors contribute in a
significant way to the total prediction. Compared with $a_e$, $\amu$ is also
much better suited to unveil or constrain ``new physics'' effects. For a
lepton $l$, their contribution to $a_l$ is generally proportional to
$m_l^2/\Lambda^2$, where $m_l$ is the mass of the lepton and $\Lambda$ is
the scale of ``new physics'', thus leading to an $(m_{\mu}/m_e)^2 \sim
4\times 10^4$ relative enhancement of the sensitivity of the muon versus the
electron anomalous magnetic moment. The anomalous magnetic moment of the
$\tau$ would thus offer the best opportunity to detect ``new physics'', but
the very short lifetime of this lepton makes such a measurement very
difficult at the moment.

In a sequence of increasingly more precise
measurements~\cite{BNL00,BNL01,BNL02,BNL04}, the E821 Collaboration at the
Brookhaven Alternating Gradient Synchrotron has reached a fabulous relative
precision of 0.5 parts per million (ppm) in the determination of $\amu$,
providing a very stringent test of the {\small SM}. Even a tiny
statistically significant discrepancy from the {\small SM} prediction could
be the harbinger for ``new physics''~\cite{CM01}.

Several excellent reviews exist on the topic presented here. Among them, I
refer the interested reader to refs.~\cite{KM90,CM99, HK99, MT00, MLR01,
Me01, Ny03, Kn03, DM04}. In this article I will provide an update and a
review of the theoretical prediction for $\amu$ in the {\small SM},
analyzing in detail the three contributions into which $\amu^{\mysmall SM}$
is usually split: {\small QED}, electroweak ({\small EW}) and hadronic. They
are respectively discussed in secs.~\ref{sec:QED}, \ref{sec:EW} and
\ref{sec:HAD}. A numerical re-evaluation of the two- and three-loop {\small
QED} contributions employing recently updated values for the lepton masses
is presented in secs.~\ref{sec:QED2} and \ref{sec:QED3}. Comparisons between
$\amu^{\mysmall SM}$ results and the current experimental determination
$\amu^{\mbox{$\scriptscriptstyle{EXP}$}}$ are given in
sec.~\ref{sec:COMP}. Conclusions are drawn in sec.~\ref{sec:CONC}.

\section{The QED Contribution to \bm $\amu$ \ubm}
\label{sec:QED}

The {\small QED} contribution to the anomalous magnetic moment of the muon
is defined as the contribution arising from the subset of {\small SM}
diagrams containing only leptons ($e,\mu,\tau$) and photons.  As a
dimensionless quantity, it can be cast in the following general
form~\cite{KM90,KNO90}
\be
    \amu^{\mysmall QED} = A_1 + A_2(m_{\mu}/m_e) + A_2(m_{\mu}/m_{\tau}) + 
                          A_3(m_{\mu}/m_e,m_{\mu}/m_{\tau}),    
\label{eq:amuqedgeneral}
\ee
where $m_e$, $m_{\mu}$ and $m_{\tau}$ are the masses of the electron, muon
and tau respectively. The term $A_1$, arising from diagrams containing only
photons and muons, is mass independent (and is therefore the same for the
{\small QED} contribution to the anomalous magnetic moment of all three
charged leptons).  In contrast, the terms $A_2$ and $A_3$ are functions of
the indicated mass ratios, and are generated by graphs containing also
electrons and taus. The renormalizability of {\small QED} guarantees that
the functions $A_i$ ($i=1,2,3$) can be expanded as power series in
$\alpha/\pi$ and computed order-by-order
\be
    A_i \,= A_i^{(2)}\left(\frac{\alpha}{\pi} \right)
    + A_i^{(4)}\left(\frac{\alpha}{\pi} \right)^{\!2}
    + A_i^{(6)}\left(\frac{\alpha}{\pi} \right)^{\!3}
    + A_i^{(8)}\left(\frac{\alpha}{\pi} \right)^{\!4}
    + A_i^{(10)}\left(\frac{\alpha}{\pi} \right)^{\!5} +\cdots.
\ee
%

\subsection{One-loop Contribution}

\begin{figure}[h]
\begin{center}
\includegraphics[width=8cm]{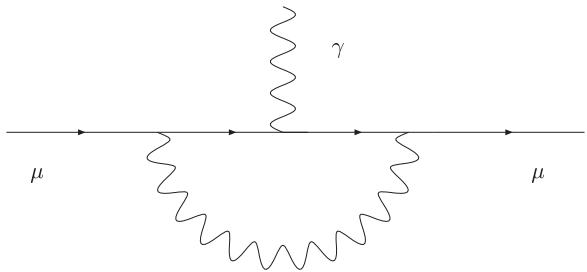}
\caption{{\sf Lowest-order {\small QED} contribution to $\amu$.}}
\label{fig:schwinger}
\end{center}
\end{figure}

Only one diagram, shown in fig.~\ref{fig:schwinger}, is involved in the
evaluation of the lowest-order contribution (second-order in the electric
charge); it provides the famous result by Schwinger~\cite{Sch48}, $A_1^{(2)}
= 1/2$ ($A_2^{(2)}=A_3^{(2)} = 0$).

\vspace{-5mm}
\subsection{Two-loop Contribution}
\label{sec:QED2}

\begin{figure}[h]
\begin{center}
\includegraphics[width=14cm]{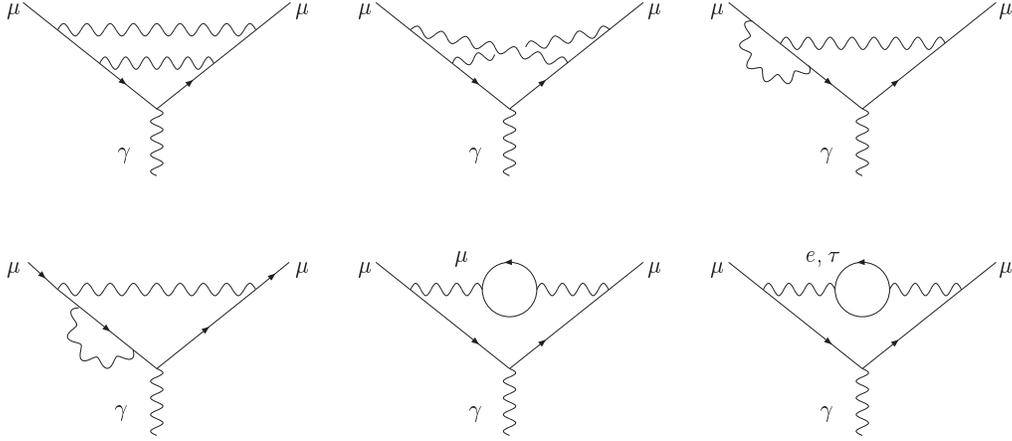}
\caption{{\sf The {\small QED} diagrams contributing to the muon $g$$-$$2$ in
         order $\alpha^2$. The mirror reflections (not shown) of the third
         and fourth diagrams must be included as well.}}
\label{fig:qed2}
\end{center}
\end{figure}

At fourth order, seven diagrams contribute to $A_1^{(4)}$, one to
$A_2^{(4)}(m_{\mu}/m_e)$ and one to $A_2^{(4)}(m_{\mu}/m_{\tau})$. They are
depicted in fig.~\ref{fig:qed2}. The coefficient $A_1^{(4)}$ has been known
for almost fifty years~\cite{So57-58,Pe57-58}: 
\be
    A_1^{(4)} = \frac{197}{144} + \frac{\pi^2}{12} 
                + \frac{3}{4}\zeta(3) - \frac{\pi^2}{2} \ln2
		= -0.328 \, 478 \, 965 \, 579 \ldots,
\label{eq:qedA14}
\ee
where $\zeta(s)$ is the Riemann zeta function of argument $s$. The
mass-dependent coefficient 
\be
     A_2^{(4)}(1/x) \,=\, \int_0^1  du \int_0^1 dv \,\,
     \frac{u^2 (1-u) v^2 (1-v^2/3)}{u^2(1-v^2) +4x^2 (1-u)},
\ee
where $x=m_l/m_{\mu}$ and $m_l$ is the mass of the virtual lepton in the
vacuum polarization subgraph, was also computed in the late
1950s~\cite{SWP57} for $m_l= m_e$ and neglecting terms of
$O(m_e/m_{\mu})$. Its exact expression was calculated in
1966~\cite{El66}. Actually, the full analytic result of~\cite{El66} can be
greatly simplified by taking advantage of the properties of the dilogarithm
${\rm Li}_2(z)=-\int_0^z dt \ln(1-t)/t$. As a result of this simplification
I obtain
\bea 
     A_2^{(4)}(1/x)   \! &=& \!
     -\frac{25}{36} - \frac{\ln x}{3} 
     +x^2 \left(4+3\ln x \right)
     +x^4 \left[ \frac{\pi^2}{3} -2\ln x \, \ln \left(\frac{1}{x}-x\right)
     -{\rm Li}_2(x^2)\right] + 
     \nonumber \\   \!&&\!
     + \, \frac{x}{2} \left(1-5 x^2\right) \!\left[\frac{\pi^2}{2} 
       - \ln x \, \ln \left( \frac{1-x}{1+x} \right) 
       - {\rm Li}_2(x) + {\rm Li}_2(-x) \right].
\label{eq:qedA24}
\eea
Note that this simple formula, contrary to the one in ref.~\cite{El66}, can
be directly used both for $0<x<1$ (the case of the electron loop) and for
$x\geq 1$ (tau loop). In the latter case, the imaginary parts developed by
the dilogarithms ${\rm Li}_2(x)$ and ${\rm Li}_2(x^2)$ are exactly canceled
by the corresponding ones arising from the logarithms. For $x=1$ (muon
loop), \eq{qedA24} gives $A_2^{(4)}(1) = 119/36 - \pi^2/3$; of course, this
contribution is already part of $A_1^{(4)}$ in \eq{qedA14}. Evaluation of
\eq{qedA24} with the latest recommended values for the muon-electron mass
ratio $m_{\mu}/m_e = 206.768\,2838\,(54)$~\cite{MT04}, and the ratio of
$m_{\mu} = 105.658\,3692\,(94)$ MeV~\cite{MT04} and $m_{\tau}=
1776.99\,(29)$ MeV~\cite{PDG04} yields
\bea
     A_2^{(4)}(m_{\mu}/m_e)      & = & 1.094\,258\,3111 \, (84)  
\label{eq:qedA24e}
\\
     A_2^{(4)}(m_{\mu}/m_{\tau}) & = & 0.000\,078\,064 \, (25),
\label{eq:qedA24tau}
\eea
where the standard uncertainties are only caused by the measurement
uncertainties of the lepton mass ratios. Eqs.~(\ref{eq:qedA24e}) and
(\ref{eq:qedA24tau}) provide the first re-evaluation of these coefficients
with the recently updated {\small CODATA} and {\small PDG} mass ratios of
refs.~\cite{MT04,PDG04}. These new values differ visibly from older ones
(see refs.~\cite{CM99, MT00}) based on previous measurements of the mass
ratios, but the change induces only a negligible shift in the total {\small
QED} prediction.  Note that the $\tau$ contribution in \eq{qedA24tau}
provides a $\sim \! 42 \times 10^{-11}$ contribution to $\amu^{\mysmall
QED}$.  As there are no two-loop diagrams containing both virtual electrons
and taus, $A_3^{(4)}(m_{\mu}/m_e,m_{\mu}/m_{\tau}) = 0$. Adding up
eqs.~(\ref{eq:qedA14}), (\ref{eq:qedA24e}) and (\ref{eq:qedA24tau}) I get
the new two-loop {\small QED} coefficient
\be
    C_2 =  A_1^{(4)} + A_2^{(4)}(m_{\mu}/m_e) + 
               A_2^{(4)}(m_{\mu}/m_{\tau}) = 0.765 \, 857 \, 410 \,(27).
\label{eq:qedC2}
\ee 
The uncertainties in $A_2^{(4)}(m_{\mu}/m_e)$ and
$A_2^{(4)}(m_{\mu}/m_{\tau})$ have been added in quadrature. The resulting
error $\delta C_2 = 2.7 \times 10^{-8}$ leads to a tiny $0.01 \times
10^{-11}$ uncertainty in $\amu^{\mysmall QED}$.

%

\subsection{Three-loop Contribution}
\label{sec:QED3}

More than one hundred diagrams are involved in the evaluation of the
three-loop (sixth-order) {\small QED} contribution. Their analytic
computation required approximately three decades, ending in the late 1990s.

\begin{figure}[h]
\begin{center}
\includegraphics[width=13cm]{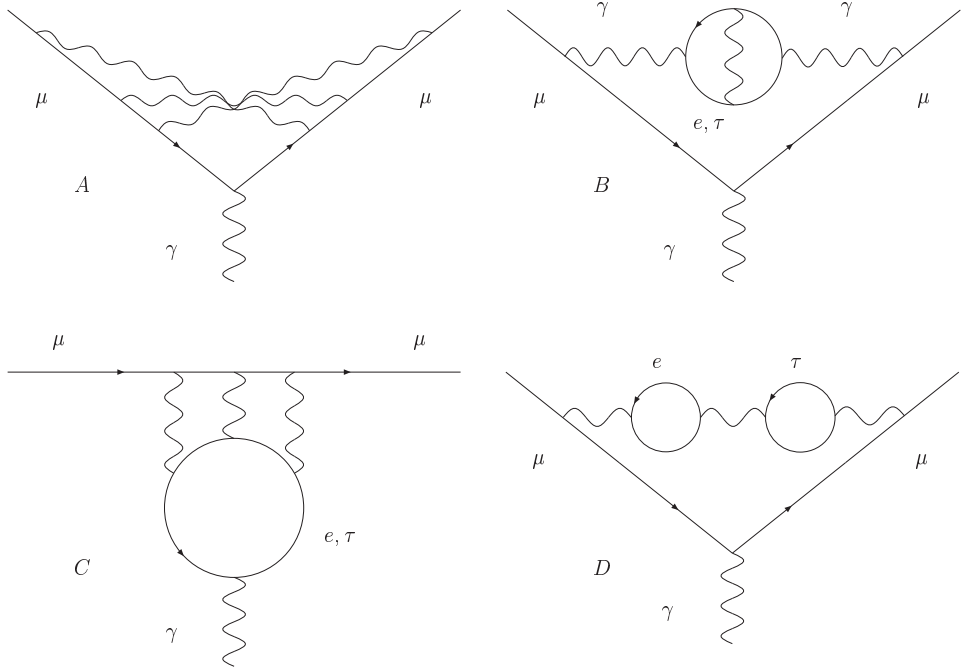}
\caption{{\sf Examples of {\small QED} diagrams contributing to the muon
         $g$$-$$2$ in order $\alpha^3$. $A$, a ``triple-cross'' diagram.
         $B$ ($C$), sixth-order muon vertex obtained by insertion of an
         electron or $\tau$ vacuum polarization (light-by-light)
         subdiagram. $D$, graph with $e$ and $\tau$ loops in the photon
         propagator.}}
\label{fig:qed3}
\end{center}
\end{figure}

The coefficient $A_1^{(6)}$ arises from 72 diagrams. Its calculation in
closed analytic form is mainly due to Remiddi and his
collaborators~\cite{Remiddi,LR96}, who completed it in 1996 with the
evaluation of the last class of diagrams, the non-planar ``triple cross''
topologies (see, for example, fig.~\ref{fig:qed3} $A$)~\cite{LR96}. The
result reads:
\bea 
     A_1^{(6)} \! &=& \!  \frac{83}{72} \pi^2 \zeta(3) - \frac{215}{24}
     \zeta(5) + \frac{100}{3} \left[a_4 + \frac{1}{24} \left( \ln^2
     2 -\pi^2 \right) \ln^2 2 \right] - \frac{239}{2160} \pi^4
     \nonumber \\ 
     \!&&\! + \frac{139}{18} \zeta(3) - \frac{298}{9} \pi^2 \ln2 +
     \frac{17101}{810} \pi^2 + \frac{28259}{5184}= 1.181 \,241\,4566 \ldots,
\label{eq:qedA16}
\eea
where $a_4=\sum_{n=1}^{\infty} 1/(2^n n^4)= {\rm Li}_4(1/2) = 0.517 \, 479\,
061 \,674 \ldots$.

The calculation of the exact expression for the coefficient $A_2^{(6)}(m/M)$
for arbitrary values of the mass ratio $(m/M)$ was completed in 1993 by
Laporta and Remiddi~\cite{La93,LR93} (earlier works include
refs.~\cite{Ki67, A26early}). For our analysis, $m=m_{\mu}$, and $M=m_e$ or
$m_{\tau}$. This coefficient can be further split into two parts: the first
one, $A_2^{(6)}(m/M,\mbox{vp})$, receives contributions from 36 diagrams
containing electron or $\tau$ vacuum polarization loops (see, for example,
fig.~\ref{fig:qed3} $B$)~\cite{La93}, whereas the second one,
$A_2^{(6)}(m/M,\mbox{lbl})$, is due to 12 light-by-light scattering diagrams
with electron or $\tau$ loops (like the graph of fig.~\ref{fig:qed3}
$C$)~\cite{LR93}. The exact expression for $A_2^{(6)}(m/M)$ in closed
analytic form is quite complicated, containing hundreds of polylogarithmic
functions up to fifth degree (for the light-by-light diagrams) and complex
arguments (for the vacuum polarization contribution). It also includes
harmonic polylogarithms~\cite{HarmPol}. As it is very lengthy, it was not
listed in the original papers~\cite{La93,LR93} which provided, however,
useful series expansions in the mass ratio $(m/M)$ for the cases of physical
relevance. Using the exact expressions in closed analytic form kindly
provided to me by the authors, and the latest values for the mass ratios
mentioned above, I obtain the following values
\bea
     A_2^{(6)}(m_{\mu}/m_e,\mbox{vp}) &=&  \:\: 1.920\, 455 \, 130 \, (33),
\label{eq:qedA26evac}
     \\
     A_2^{(6)}(m_{\mu}/m_e,\mbox{lbl}) &=&      20.947 \, 924 \, 89\,(16),
\label{eq:qedA26elbl}
     \\
     A_2^{(6)}(m_{\mu}/m_{\tau},\mbox{vp})&=& \!\! -0.001\,782\,33 \, (48),  
\label{eq:qedA26tauvac}
     \\
     A_2^{(6)}(m_{\mu}/m_{\tau},\mbox{lbl})&=&\,\:\,0.002\,142\,83\, (69).
\label{eq:qedA26taulbl}
\eea
The sums of eqs.~(\ref{eq:qedA26evac})--(\ref{eq:qedA26elbl}) and 
eqs.~(\ref{eq:qedA26tauvac})--(\ref{eq:qedA26taulbl}) are
\bea
     A_2^{(6)}(m_{\mu}/m_e) &=&      22.868 \, 380 \, 02\,(20),
\label{eq:qedA26e}
     \\
     A_2^{(6)}(m_{\mu}/m_{\tau})&=&\,\:\,0.000\, 360 \, 51\, (21);
\label{eq:qedA26tau}
\eea
to determine the uncertainties, the correlation of the addends has been
taken into account. Eqs.~(\ref{eq:qedA26evac})--(\ref{eq:qedA26tau}) provide
the first re-evaluation of these coefficients with the recently updated
{\small CODATA} and {\small PDG} mass ratios of
refs.~\cite{MT04,PDG04}. These new values differ visibly from older ones
(see refs.~\cite{CM99, MT00}) based on previous measurements of the mass
ratios, but the change induces only a negligible shift in the total {\small
QED} prediction.  Note the large contribution from the electron
light-by-light diagrams, \eq{qedA26elbl} -- its leading term is $(2/3)\pi^2
\ln(m_{\mu}/m_e)$~\cite{LS77}. More generally, it was shown in~\cite{Ye89}
that the $O(\alpha^{2n+1})$ contribution to $\amu^{\mysmall QED}$, from
diagrams in which the electron light-by-light subgraph is connected with
$2n+1$ photons to the muon, contains a large $\pi^{2n}\ln(m_{\mu}/m_e)$ term
with a coefficient of $O(1)$.

The analytic calculation of the three-loop diagrams with both electron and
$\tau$ loop insertions in the photon propagator (see fig.~\ref{fig:qed3}
$D$) became available in 1999~\cite{CS99}. This analytic result yields the
numerical value
\be
     A_3^{(6)}(m_{\mu}/m_e,m_{\mu}/m_{\tau}) 
     \,=\, 0.000 \, 527 \, 66 \, (17),
\label{eq:qedA36}
\ee
providing a small $0.7 \times 10^{-11}$ contribution to $\amu^{\mysmall
QED}$. The error, $1.7 \times 10^{-7}$, is caused by the uncertainty of the
ratio $m_{\mu}/m_{\tau}$. Combining the three-loop results presented above,
I obtain the new sixth-order {\small QED} coefficient
\bea
    C_3 &=&  A_1^{(6)} + 
               A_2^{(6)}(m_{\mu}/m_e) + 
               A_2^{(6)}(m_{\mu}/m_{\tau}) +
	       A_3^{(6)}(m_{\mu}/m_e,m_{\mu}/m_{\tau}) 
	       \nonumber \\
	&=&       24.050 \, 509 \,64 \, (43).
\label{eq:qedC3}
\eea 
The error $\delta C_3 = 4.3 \times 10^{-7}$, due to the measurement
uncertainties of the lepton masses, has been determined considering the
correlation of the addends. It induces a negligible $O(10^{-14})$
uncertainty in $\amu^{\mysmall QED}$.

In parallel to these analytic results, numerical methods were also
developed, mainly by Kinoshita and his collaborators, for the evaluation of
the full set of three-loop diagrams~\cite{Ki90, KM90}.
%
%
\subsection{Four-loop Contribution}

More than one thousand diagrams enter the evaluation of the four-loop
{\small QED} contribution to $\amu$. As only few of them are known
analytically~\cite{A8analytic}, this eighth-order term has thus far been
evaluated only numerically. This formidable task was first accomplished by
Kinoshita and his collaborators in the early 1980s~\cite{KL81,KNO84}. Since
then, they made a major effort to continuously improve this
result~\cite{KM90, HK99, KNO90, KL83-89, Ki93, KN03}, also benefiting from
fast advances in computing power. The latest analysis appeared in
ref.~\cite{KN04}. One should realize that this eighth-order {\small QED}
contribution, being about six times larger than the present experimental
uncertainty of $\amu$, is crucial for the comparison between the {\small SM}
prediction of $\amu$ and its experimental determination.

There are 891 four-loop diagrams contributing to the mass-independent
coefficient $A_1^{(8)}$. Its latest published value is $A_1^{(8)} = -1.7502
\, (384)$~\cite{KN03}, where the error is caused by the numerical
procedure. This coefficient has undergone a small revision in
ref.~\cite{KN03}. In September 2004 Kinoshita reported a new preliminary
updated result~\cite{Ki04},
\be
    A_1^{(8)} \, = \, -1.7093 \, (42).
\label{eq:qedA18}
\ee
Note the small shift in the central value and the significant reduction of
the numerical uncertainty of this new result. I will adopt it for the value
of $A_1^{(8)}$. The latest value of the coefficient
$A_2^{(8)}(m_{\mu}/m_e)$, arising from 469 diagrams, is~\cite{KN04}
\be
    A_2^{(8)}(m_{\mu}/m_e) \,=\, 132.6823 \, (72).
\label{eq:qedA28e}
\ee
This value is significantly higher than the older one, $127.50
\,(41)$~\cite{HK99} (its precision is impressively higher too) shifting up
the value of $\amu^{\mysmall QED}$ by a non-negligible $\sim \! 15 \times
10^{-11}$. This difference is partly accounted for by the correction of a
program error described in ref.~\cite{KN03}, but is mostly due to the fact
that the computation of the older value suffered from insufficient numerical
precision.  The term $A_2^{(8)}(m_{\mu}/m_{\tau})$ has been roughly
estimated to give an $O(10^{-13})$ contribution to $\amu^{\mysmall QED}$ --
it can be safely ignored for now~\cite{KN04}. The numerical evaluation of
the 102 diagrams containing both electron and $\tau$ loop insertions yields
the three-mass coefficient~\cite{KN04}
\be
    A_3^{(8)}(m_{\mu}/m_e,m_{\mu}/m_{\tau})  
     \,=\, 0.037 \,594 \, (83),
\label{eq:qedA28tau}
\ee
which provides a small $O(10^{-12})$ contribution to $\amu^{\mysmall QED}$.
Adding up the four-loop results described above, we obtain the eighth-order
{\small QED} coefficient
\be
    C_4 \, \simeq \,   A_1^{(8)} + 
               A_2^{(8)}(m_{\mu}/m_e) + 
	       A_3^{(8)}(m_{\mu}/m_e,m_{\mu}/m_{\tau}) 
	\,=\,  131.011 \,(8).
\label{eq:qedC4}
\ee 
Note that this expression does not contain the term
$A_2^{(8)}(m_{\mu}/m_{\tau})$, which has been roughly estimated to be of the
same order of magnitude of the uncertainty on the r.h.s.\ of
\eq{qedC4}. However, this uncertainty, $0.008$, causes only a tiny $0.02
\times 10^{-11}$ error in $\amu^{\mysmall QED}$.

\subsection{Five-loop Contribution}

The evaluation of the five-loop {\small QED} contribution is in
progress~\cite{Ki04}. The existing estimates are mainly based on the
experience accumulated computing the sixth- and eighth-order terms, and
include only specific contributions enhanced by powers of $\ln(m_{\mu}/m_e)$
times powers of $\pi$. The first estimate, $C_5 = 570 \,(140)$, provided by
Kinoshita and collaborators in 1990~\cite{KNO90}, considered the
contribution of graphs containing an electron light-by-light subdiagram with
one-loop vacuum polarization insertions. A few other predictions for $C_5$
exist, and classes of diagrams were computed or estimated with various
methods~\cite{Ye89, MY89, Ka92, Br93, Ka93, La94, EKS94, KS94}.  In
September 2004 Kinoshita reported a new very preliminary result~\cite{Ki04},
\be C_5 \, \simeq \, A_2^{(10)}(m_{\mu}/m_e) \, = \, 677 \, (40),
\label{eq:qedC5}
\ee 
(9080 diagrams contribute to $A_2^{(10)}(m_{\mu}/m_e)$!) corresponding to a
$4.6\,(0.3) \times 10^{-11}$ contribution to $\amu^{\mysmall QED}$. This is
the value of $C_5$ I will employ. The uncertainty in this new estimate of
the tenth-order term ($0.3 \times 10^{-11}$) no longer dominates the error
of the total {\small QED} prediction (see next section). Efforts to improve
upon the evaluation of $C_5$ are presently being pursued by Kinoshita and
Nio.

\subsection{The Numerical Value of {\bm $\amu^{\mysmall QED}$\ubm} }

Adding up all the above contributions and using the latest {\small CODATA}
recommended value for the fine-structure constant~\cite{MT04}, known to 3.3
ppb,
\be
     \alpha^{-1} \, = \,  137.035 \,999 \,11 \,(46),    
\label{eq:alphaMT04}
\ee 
I obtain the following value for the {\small QED} contribution to the muon
$g$$-$$2$:
\be
    \amu^{\mysmall QED} = 
    116 \, 584 \, 718.8 \, (0.3)\,(0.4)  \times 10^{-11}.       
\label{eq:qed}
\ee
The first error is due to the uncertainties of the $O(\alpha^2)$,
$O(\alpha^4)$ and $O(\alpha^5)$ terms, and is strongly dominated by the last
of them.  (The uncertainty of the $O(\alpha^3)$ term is negligible.)  The
second error is caused by the 3.3 ppb uncertainty of the fine-structure
constant $\alpha$.  When combined in quadrature, these uncertainties yield
$\delta \amu^{\mysmall QED} = 0.5 \times 10^{-11}$. The value of
$\amu^{\mysmall QED}$ in \eq{qed} is close to that presented by Kinoshita
in~\cite{Ki04}, $\amu^{\mysmall QED} = 116 \, 584 \, 717.9 \,(0.3)\,(0.9)
\times 10^{-11}$, and has a smaller error. This latter result was in fact
derived using the value of $\alpha$ determined from atom interferometry
measurements~\cite{alpha}, $\alpha^{-1}=137.036\,000\,3\,(10)$ (7.3 ppb),
which has a larger uncertainty than the latest {\small CODATA} value
employed for \eq{qed}.

\section{The Electroweak Contribution}
\label{sec:EW}

The electroweak ({\small EW}) contribution to the anomalous magnetic moment
of the muon is suppressed by a factor $(m_{\mu}/\mw)^2$ with respect to the
{\small QED} effects. The one-loop part was computed in 1972 by several
authors~\cite{ew1loop}. Back then, the experimental uncertainty of $\amu$
was one or two orders of magnitude larger than this one-loop contribution.
Today it's less than one-third as large.

\subsection{One-loop Contribution}

\begin{figure}[h]
\begin{center}
\includegraphics[width=14cm]{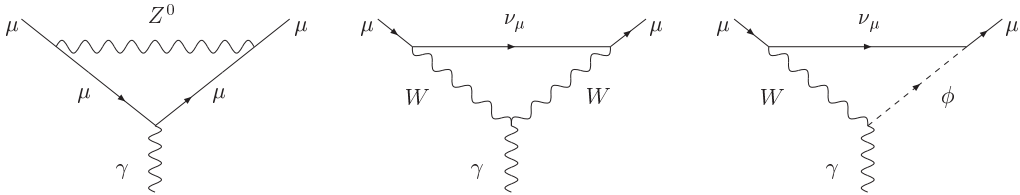}
\caption{{\sf One-loop electroweak contributions to $a_{\mu}$. The diagram
  with a W and a Goldstone boson ($\phi$) must be counted twice. The
  diagrams with the Higgs boson loop and with two Goldstone boson couplings
  to the muon are suppressed by a factor $m^2_{\mu}/M^2_{\mysmall{Z,W,H}}$ and
  are not drawn.}}
\label{fig:ew1}
\end{center}
\end{figure}

The analytic expression for the one-loop {\small EW} contribution to $\amu$,
due to the diagrams in fig.~\ref{fig:ew1}, reads
\be
     \amu^{\mysmall EW} (\mbox{1 loop}) =
     \frac{5 G_{\mu} m^2_{\mu}}{24 \sqrt{2} \pi^2}
     \left[ 1+ \frac{1}{5}\left(1-4\sin^2\!\theta_{\mysmall{W}}\right)^2 
       + O \left( \frac{m^2_{\mu}}{M^2_{\mysmall{Z,W,H}}} \right) \right],
\label{eq:EWoneloop}
\ee
where $G_{\mu}=1.16637(1) \times 10^{-5}\gev^{-2}$ is the Fermi coupling
constant, $\mz$, $\mw$ and $\mh$ are the masses of the $Z$, $W$ and Higgs
bosons, and $\theta_{\mysmall{W}}$ is the weak mixing angle.  Closed
analytic expressions for $\amu^{\mysmall EW} (\mbox{1 loop})$ taking exactly
into account the $m^2_{\mu}/M^2_{\mysmall{B}}$ dependence ($B=Z,W,$ Higgs,
or other hypothetical bosons) can be found in refs.~\cite{Studenikin}.
Following~\cite{CMV03}, I employ for $\sin^2\!\theta_{\mysmall{W}}$ the
on-shell definition $\sin^2\!\theta_{\mysmall{W}} =
1-M^2_{\mysmall{W}}/M^2_{\mysmall{Z}}$~\cite{Si80}, where
$\mz=91.1875(21)\gev$ and $\mw$ is the theoretical {\small SM} prediction of
the $W$ mass. The latter can be easily derived from the simple analytic
formulae of ref.~\cite{FOPS},
\be
      \mw = \left[ 80.4077 
	- 0.05738 \, \ln \! \left(\frac{\mh}{100\gev}\right) -
	0.00892\, \ln^2 \! 
	  \left(\frac{\mh}{100\gev}\right)\right]\!\!\gev,
\label{eq:fops}
\ee
(on-shell scheme {\small II} with $\Delta \alpha_h^{(5)}=0.02761 \,(36)$,
$\alpha_s(\mz)=0.118 \,(2)$ and $M_{\rm\scriptstyle top}=$ $178.0 \, (4.3)$ GeV
\cite{newTOP}), leading to $\mw =80.383\gev$ for $\mh=150\gev$, compared
with the direct experimental value $\mw=80.425 \,(38)\gev$ \cite{PDG04}, which
corresponds to a small $\mh$~\cite{FOS04}. For $\mh=150\gev$,
eq.~(\ref{eq:EWoneloop}) thus gives
\be
    \amu^{\mysmall EW} (\mbox{1 loop}) = 194.8 \times 10^{-11},
\label{eq:EWoneloopNumber}
\ee
but this value encompasses the predictions derived from a wide range of
values of $\mh$ varying from 114.4 GeV, the current lower bound at 95\%
confidence level~\cite{LEPHIGGS}, up to a few hundred GeV.

The contribution of the Higgs diagram alone, part of the
$O(m^2_{\mu}/M^2_{\mysmall{Z,W,H}})$ terms of eq.~(\ref{eq:EWoneloop}),
is~\cite{KM90,Studenikin}
\be
     \amu^{\mysmall EW,H} (\mbox{1 loop}) = 
     \frac{G_{\mu} m^2_{\mu}}{4 \sqrt{2} \pi^2} 
     \left[\frac{\log R_{\mysmall{H}}}{R_{\mysmall{H}}}
	   -\frac{7}{6R_{\mysmall{H}}} 
	     + O \left( \frac{1}{R_{\mysmall{H}^2}} \right) \right],
\ee
where $R_{\mysmall{H}}=M^2_{\mysmall{H}}/m^2_{\mu}$.  Given the current
lower bound $M_{\mysmall{H}} > 114.4$ GeV (95\% {\small CL}),
$\amu^{\mysmall EW,H} (\mbox{1 loop})$ is smaller than $3 \times 10^{-14}$
and can be safely neglected.

\subsection{Higher-order Contributions}

The two-loop {\small EW} contribution to $\amu$ was computed in 1995 by
Czarnecki, Krause and Marciano~\cite{CKM95a,CKM95b}. This remarkable
calculation, probably the first (and still one of the very few) complete
two-loop electroweak computation, leads to a significant reduction of the
one-loop prediction. Na\"{\i}vely one would expect the two-loop {\small EW}
contribution $\amu^{\mysmall EW} (\mbox{2 loop})$ to be of order
$(\alpha/\pi) \times \amu^{\mysmall EW} (\mbox{1 loop})$, and thus
negligible, but this turns out not to be so. As first noticed in
1992~\cite{KKSS}, $\amu^{\mysmall EW} (\mbox{2 loop})$ is actually quite
substantial because of the appearance of terms enhanced by a factor of
$\ln(M_{\mysmall{Z,W}}/m_f)$, where $m_f$ is a fermion mass scale much
smaller than $\mw$.

The two-loop contributions to $\amu^{\mysmall EW}$ can be divided into
fermionic and bosonic parts; the former includes all two-loop {\small EW}
corrections containing closed fermion loops, whereas all other contributions
are grouped into the latter. The full two-loop calculation involves 1678
diagrams in the linear 't Hooft-Feynman gauge~\cite{Kaneko95}. As a check,
the authors of~\cite{CKM95a,CKM95b} employed both this gauge and a nonlinear
one in which the vertex of the photon, the $W$ and the unphysical charged
scalar vanishes. Their result for $\mh=150\gev$ (obtained in the
approximation $\mh \gg M_{\mysmall{W,Z}}$ computing the first two terms in
the expansion in $M^2_{\mysmall{W,Z}}/\mh^2$) was $\amu^{\mysmall EW}
(\mbox{2 loop})= -42.3(2.0)(1.8)\times 10^{-11}$, leading to a significant
reduction of $\amu^{\mysmall EW}$. The first error is meant to roughly
reflect low momentum hadronic uncertainties (more below), whereas the second
allows for a range of $\mh$ values from 114 GeV to about 250 GeV.  Note that
the contribution from $\gamma$--$Z$ mixing diagrams is not included in this
result: as it is suppressed by ($1-4\sin^2\!\theta_{\mysmall{W}}) \sim 0.1$
for quarks and ($1-4\sin^2\!\theta_{\mysmall{W}})^2$ for leptons, it was
neglected in this early calculation. It was later studied in
ref.~\cite{CMV03}: together with small contributions proportional to
$(1-4\sin^2\!\theta_{\mysmall{W}})(m_t^2/\mw^2)$ induced by the
renormalization of $\sin^2\!\theta_{\mysmall{W}}$, it shifts the above value
of $\amu^{\mysmall EW}$ down by a tiny $0.4\times 10^{-11}$.

\begin{figure}[h]
\begin{center}
\includegraphics[width=12cm]{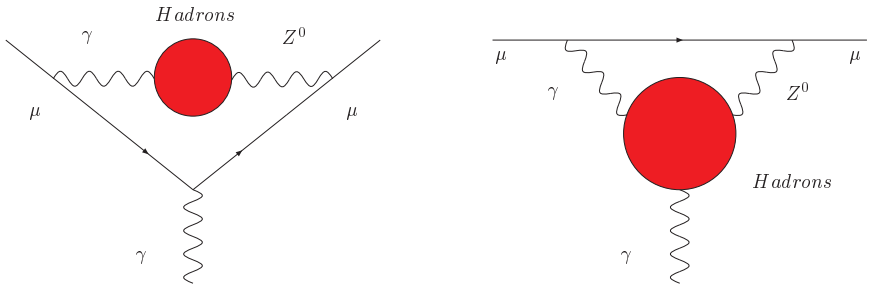}
\caption{{\sf Hadronic loops in two-loop {\small EW} contributions.}}
\label{fig:ew2}
\end{center}
\end{figure}

The hadronic uncertainties, above estimated to be $\sim 2\times 10^{-11}$,
arise from two types of two-loop diagrams: hadronic photon--$Z$ mixing, and
quark triangle loops with the external photon, a virtual photon and a $Z$
attached to them (see fig.~\ref{fig:ew2}).  
The tiny hadronic $\gamma$--$Z$ mixing terms can be evaluated either in the
free quark approximation or via a dispersion relation using data from
$e^+e^-$ annihilation into hadrons; the difference was shown to be
numerically insignificant~\cite{CMV03}. The contribution from the second
type of diagrams (the quark triangle ones), calculated in \cite{CKM95a} in
the free quark approximation, is numerically more important. The question of
how to treat properly the contribution of the light quarks was originally
addressed in ref.~\cite{PPD95} within a low-energy effective field theory
approach and was further investigated in the detailed analyses of
refs.~\cite{CMV03,KPPD02,CMV03b}. These refinements significantly improved
the reliability of the fermionic part of $\amu^{\mysmall EW}(\mbox{two
loop})$ and increased it by $2\times 10^{-11}$ relative to the free quark
calculation, leading, for $\mh=150\gev$, to~\cite{CMV03}
\be
    \amu^{\mysmall EW} = 154(1)(2)\times 10^{-11},
\label{eq:ew}
\ee
where the first error corresponds to hadronic loop uncertainties and the
second to an allowed Higgs mass range of $114\gev < \mh < 250\gev$, the
current top mass uncertainty\footnote{Indeed, although the result in
eq.~(\ref{eq:ew}) was computed for $m_t=174.3\gev$, I checked that the shift
induced in $\amu^{\mysmall EW}$ by the latest experimental value $m_t=178.0
\, (4.3)\gev$~\cite{newTOP} is well within the quoted error.} and unknown
three-loop effects.

The leading-logarithm three-loop contribution to $\amu^{\mysmall EW}$ was
first studied via a renormalization group analysis in ref.~\cite{DGi98}.
Such an analysis was revisited and refined in~\cite{CMV03}, where this
contribution was found to be extremely small (indeed, consistent with zero
to a level of accuracy of $10^{-12}$). An uncertainty of $0.2\times
10^{-11}$, included in eq.~(\ref{eq:ew}), has been conservatively
assigned to $\amu^{\mysmall EW}$ for uncalculated three-loop
nonleading-logarithm terms.

Lastly, I would like to point out that until recently only one evaluation
existed of the two-loop bosonic part of $\amu^{\mysmall EW}$, ie,
ref.~\cite{CKM95b}. The recent calculation of ref.~\cite{HSW04}, performed
without the approximation of large Higgs mass previously employed, agrees
with the result of~\cite{CKM95b}. Work is also in progress for an
independent recalculation based on the numerical methods of
refs.~\cite{Topside}.

\section{The Hadronic Contribution}
\label{sec:HAD}

In this section I will analyze the contribution to the muon $g$$-$$2$
arising from {\small QED} diagrams involving hadrons. Hadronic effects in
(two-loop) {\small EW} contributions are already included in $\amu^{\mysmall
EW}$ (see the previous section).

\subsection{Leading-order Hadronic Contribution}
\label{subsec:HLO}

\begin{figure}[h]
\begin{center}
\includegraphics[width=7cm]{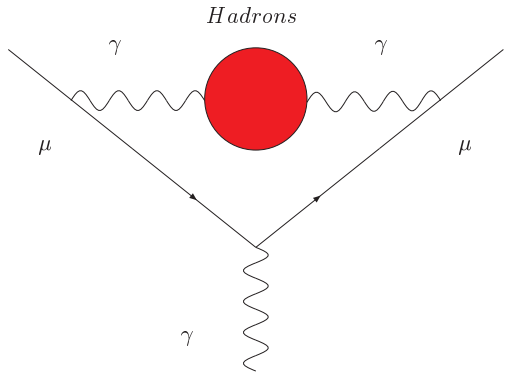}
\caption{{\sf Leading hadronic contribution to $\amu$.}}
\label{fig:hlo}
\end{center}
\end{figure}

The leading hadronic contribution to the muon $g$$-$$2$,
$\amu^{\mbox{$\scriptscriptstyle{HLO}$}}$, is due to the hadronic vacuum
polarization correction to the internal photon propagator of the one-loop
diagram (see diagram in fig.~\ref{fig:hlo}). 
The evaluation of this $O(\alpha^2)$ diagram involves long-distance {\small
QCD} for which perturbation theory cannot be employed. However, using
analyticity and unitarity (the optical theorem), Bouchiat and
Michel~\cite{BM61} showed long ago that this contribution can be computed
from hadronic $e^+ e^-$ annihilation data via the dispersion
integral~\cite{BM61,LBdRGdR}
\be
      \amu^{\mbox{$\scriptscriptstyle{HLO}$}}= 
      \frac{1}{4\pi^3} 
      \int^{\infty}_{4m^2_\pi} ds \, K(s) \, \sigma^{(0)}(s) =
      \frac{\alpha^2}{3\pi^2}
       \int^{\infty}_{4m^2_\pi} \frac{ds}{s} \, K(s) \, R(s),
\label{eq:hlo}
\ee
where $\sigma^{(0)}(s)$ is the experimental total cross section for $e^+
e^-$ annihilation into any hadronic state, with extraneous {\small QED}
radiative corrections subtracted off (more later), and $R(s)$ is the ratio
of $\sigma^{(0)}(s)$ and the high-energy limit of the Born cross section for
$\mu$-pair production: $R(s) = \sigma^{(0)}(s)/(4\pi \alpha^2\!/3s)$.  The
kernel $K(s)$ is the well-known function
\be
                K(s)= \int_0^1 \!dx \frac{x^2 (1-x)}
                {x^2 +(1-x)s/m_\mu^2}
\ee
(see ref.~\cite{EJ95} for some of its explicit representations and their
suitability for numerical evaluations). It decreases monotonically for
increasing $s$, and for large $s$ it behaves as $m_\mu^2/3s$ to a good
approximation. For this reason the low-energy region of the dispersive
integral is enhanced by $\sim 1/s^2$. About 91\% of the total contribution
to $\amu^{\mbox{$\scriptscriptstyle{HLO}$}}$ is accumulated at
center-of-mass energies $\sqrt{s}$ below 1.8 GeV and 73\% of
$\amu^{\mbox{$\scriptscriptstyle{HLO}$}}$ is covered by the two-pion final
state which is dominated by the $\rho(770)$ resonance~\cite{DEHZ03}.
Exclusive low-energy $e^+e^-$ cross sections have been mainly measured by
experiments running at $e^+e^-$ colliders in Novosibirsk ({\small OLYA, TOF,
ND, CMD, CMD-2, SND}) and Orsay ({\small M3N, DM1, DM2}), while at higher
energies the total cross section ratio $R(s)$ has been measured inclusively
by the experiments $\gamma \gamma 2$, {\small MARK I, DELCO, DASP, PLUTO,
LENA}, Crystal Ball, {\small MD-1, CELLO, JADE, MARK-J, TASSO, CLEO, CUSB,
MAC}, and {\small BES}.  Perturbative {\small QCD} becomes applicable at
higher loop momenta, so that at some energy scale one can switch from data
to {\small QCD}~\cite{pQCD,DH98a,DH98b}.

Detailed evaluations of the dispersive integral in eq.~(\ref{eq:hlo}) have
been carried out by several authors~\cite{EJ95, DEHZ03, DH98a, DH98b,
B85,KNO85,CLY85,MD89, AY95, BW96, ADH98, SN01, JatSirlin, dTY01, CLS01,
DEHZ02, HMNT02, J03, ELZ04, ELZ03, HMNT03, dTY04, DEHZ04}. A prominent role
among all data sets is played by the precise measurements by the {\small
CMD-2} detector at the {\small VEPP-2M} collider in
Novosibirsk~\cite{CMD2-99,CMD2-01,CMD2-03} of the cross section for
$e^+e^-\rightarrow \pi^+\pi^-$ at values of $\sqrt s$ between 0.61 and 0.96
GeV (ie, $s \in [0.37,0.93]\,{\rm GeV}^2$). The quoted systematic error of
these data is 0.6\%~\cite{CMD2-03}, dominated by the uncertainties in the
radiative corrections (0.4\%). In July 2004, also the {\small KLOE}
experiment at the {\small DA$\Phi$NE} collider in Frascati presented the
final analysis~\cite{KLOE-04} of the 2001 data for the precise measurement
of $\sigma(e^+e^-\rightarrow \pi^+\pi^-)$ via the radiative return
method~\cite{RadRet} from the $\phi$ resonance. In this case the machine is
operating at a fixed center-of-mass energy $W \simeq$ 1.02 GeV, the mass of
the $\phi$ meson, and initial-state radiation is used to reduce the
invariant mass of the $\pi^+\pi^-$ system. In~\cite{KLOE-04} the cross
section $\sigma(e^+e^-\rightarrow \pi^+\pi^-)$ was extracted for the range
$s \in [0.35,0.95]\,{\rm GeV}^2$ with a systematic error of 1.3\% (0.9\%
experimental and 0.9\% theoretical) and a negligible statistical one.  The
study of the $e^+e^-\rightarrow \pi^+\pi^-$ process via the initial-state
radiation method is also in progress at the {\small BABAR} detector at the
{\small PEP-II} collider in {\small SLAC}~\cite{BABAR}. This analysis will
be important to further assess the consistency of the $e^+e^-$ data.  The
{\small BABAR} collaboration has already presented data for the $\pi^+ \pi^-
\pi^0$ final state~\cite{BABAR+-0}, and preliminary ones for the process
$e^+e^- \rightarrow 2\pi^+ 2\pi^-$~\cite{BABAR}. On the theoretical side,
the properties of analyticity, unitarity and chiral symmetry provide strong
constraints for the pion form factor $F_\pi(s)$ in the low-energy
region~\cite{GM91, BERN01,Le02,Co03,VLC,dTY04}. They can lead to further
improvements. Perhaps, also lattice {\small QCD} computations of
$\amu^{\mbox{$\scriptscriptstyle{HLO}$}}$, although not yet competitive with
the precise results of the dispersive method, may eventually rival that
precision~\cite{LATTICE}.

The hadronic contribution $\amu^{\mbox{$\scriptscriptstyle{HLO}$}}$ is of
order $7000\times 10^{-11}$. Of course, this is a small fraction of the
total {\small SM} prediction for $\amu$, but is very large compared with the
current experimental uncertainty $\delta
a_{\mu}^{\mbox{$\scriptscriptstyle{EXP}$}} = 60\times 10^{-11}$. Indeed, as
$\delta a_{\mu}^{\mbox{$\scriptscriptstyle{EXP}$}}$ is less than one percent
of $\amu^{\mbox{$\scriptscriptstyle{HLO}$}}$, precision analyses of this
hadronic term as well as full treatment of its higher-order corrections are
clearly warranted. Normally, the ``bare'' cross section $\sigma^{(0)}(s)$ is
used in the evaluation of the dispersive integral and the higher-order
hadronic corrections (see sec.~\ref{subsec:HHO}) are addressed
separately. But what does ``bare'' really mean?

The extraction of $\sigma^{(0)}(s)$ from the observed hadronic cross section
$\sigma(s)$ requires the subtraction of several radiative corrections
({\small RC}) which, at the level of precision we are aiming at, have a
substantial impact on the result.  To start with, {\small RC} must be
applied to the luminosity determination, which is based on large-angle
Bhabha scattering and muon-pair production in low-energy experiments, and
small-angle Bhabha scattering at high energies. The first step to derive
$\sigma^{(0)}(s)$ consists then in subtracting the initial-state radiative
({\small ISR}) corrections (virtual and real, described by pure {\small
QED}) from $\sigma(s)$. The resulting cross section still contains the
effects of the photon vacuum polarization corrections ({\small VP}), which
can be simply undressed by multiplying it by $\alpha^2/\alpha(s)^2$, where
$\alpha(s)$ is the effective running coupling (obviously depending on
nonperturbative contributions itself). The problem with data from old
experiments is that it's difficult to find out if (and which of) these
corrections have been included (see ref.~\cite{DEHZ02}).  The latest
analysis from {\small CMD-2}~\cite{CMD2-03} is explicitly corrected for both
{\small ISR} and {\small VP} (leptonic as well as hadronic) effects, whereas
the preliminary data~\cite{CMD2-99} of the same experiment were only
corrected for {\small ISR}. For a thorough analysis of these problems, I
refer the reader to~\cite{DEHZ02,HMNT03,HGJ02} and references therein.

All hadronic final states should be incorporated in the hadronic
contribution to the muon $g$$-$$2$, in particular final states including
photons. These final-state radiation ({\small FSR}) effects, although of
higher order ($\alpha^3$), are normally included in the leading-order
hadronic contribution $\amu^{\mbox{$\scriptscriptstyle{HLO}$}}$. I will
stick to this time-honored convention. The precise {\small CMD-2} data for
the cross section $e^+e^- \rightarrow \pi^+ \pi^-$ (quoted systematic error
of 0.6\% dominated by the uncertainties in the {\small RC}) are corrected
for {\small FSR} effects using {\em scalar} {\small QED}. I find this
worrisome. The following is done: their experimental analysis imposes cuts
to isolate the two-pion final states. These cuts exclude a large fraction of
the $\pi^+ \pi^- \gamma$ states, in particular those where the photon is
radiated off at a relatively large angle~\cite{Me01}. The fraction left is
then removed using the Monte Carlo simulation based on point-like
pions. Finally, the full {\small FSR} contribution is added {\em back} using
an analytic expression computed in scalar {\small QED} for point-like
pions~\cite{Sch89}, shifting up the value of
$\amu^{\mbox{$\scriptscriptstyle{HLO}$}}$ by $\sim 50 \times
10^{-11}$~\cite{Me01, dTY01, dTY04, HGJ02, CGKR03, DKMPS04}. (This full
scalar-{\small QED} {\small FSR} contribution is also added to older
$\pi^+ \pi^-$ data.)  This procedure is less than perfect, as it introduces
a model dependence which could be avoided by a direct measurement of the
cross section into hadronic states inclusive of photons. Any calculation
that invokes scalar {\small QED} probably falls short of what is needed.

The 2001 final analysis~\cite{CMD2-01} of the precise {\small CMD-2} $\pi^+
\pi^-$ data taken in 1994--95 substantially differed from the preliminary
one~\cite{CMD2-99} released two years earlier (based on the same data
sample). The difference mostly consisted in the treatment of {\small RC},
resulting in a reduction of the cross section by about 1\% below the $\rho$
peak and 5\% above. A second significant change occurred during the summer
of 2003, when the {\small CMD-2} collaboration discovered an error in the
Monte Carlo program for Bhabha scattering that was used to determine the
luminosity~\cite{CMD2-03}.  As a result, the luminosity was overestimated by
2--3\%, depending on energy. (Another problem was found in the {\small RC}
for $\mu$-pairs production.) Overall, the pion-pair cross section
increased by 2.1--3.8\% in the measured energy range~\cite{DEHZ03}, a
non-negligible shift. The 2004 results of the {\small KLOE} collaboration,
obtained via the radiative return method from the $\phi$ resonance, are in
fair agreement with the latest energy scan data from {\small
CMD-2}~\cite{DEHZ04, CMD2-03,KLOE-04}. Here I will only report the
evaluations of the dispersive integral in eq.~(\ref{eq:hlo}) based on the
latest {\small CMD-2} reanalysis, as it supersedes all earlier ones.  These
evaluations are in very good agreement:\footnote{I have translated the
results of ref.~\cite{dTY04} into the notation of the present article.}
\bea
\label{eq:DEHZ04}
      \mbox{\cite{DEHZ04}} \qquad\qquad&
      \amu^{\mbox{$\scriptscriptstyle{HLO}$}} \,\,= & 
      6934 \, (53)_{exp} (35)_{rad} \times 10^{-11},  \\
\label{eq:J03}
      \mbox{\cite{J03}}    \qquad\qquad&
      \amu^{\mbox{$\scriptscriptstyle{HLO}$}} \,\,= & 
      6948 \, (86) \times 10^{-11},  \\
\label{eq:ELZ04}
      \mbox{\cite{ELZ04}}  \qquad\qquad&   
       \amu^{\mbox{$\scriptscriptstyle{HLO}$}}\,\,= & 
      6934 \, (92) \times 10^{-11}, \\
\label{eq:HMNT03}
      \mbox{\cite{HMNT03}} \qquad\qquad&
      \amu^{\mbox{$\scriptscriptstyle{HLO}$}} \,\,= & 
      6924 \, (59)_{exp} (24)_{rad} \times 10^{-11},  \\
\label{eq:dTY04}
      \mbox{\cite{dTY04}}  \qquad\qquad&
      \amu^{\mbox{$\scriptscriptstyle{HLO}$}} \,\,= & 
      6944 \, (48)_{exp} (10)_{rad} \times 10^{-11}.  
\eea
The preliminary result in eq.~(\ref{eq:DEHZ04}) already includes {\small
KLOE}'s 2004 data analysis and updates the one of ref.~\cite{DEHZ03},
shifting it down by $29 \times 10^{-11}$; two thirds of this shift are due
to the inclusion of {\small KLOE}'s data. The preliminary new result in
eq.~(\ref{eq:ELZ04}) updates the value
$\amu^{\mbox{$\scriptscriptstyle{HLO}$}}= 6996 \,(85)_{exp} (19)_{rad}
(20)_{proc} \times 10^{-11}$ previously obtained by the same
authors~\cite{ELZ03}. Their central value decreased because of an
improvement of their integration procedure.

The authors of ref.~\cite{ADH98} pioneered the idea of using vector spectral
functions derived from the study of hadronic $\tau$ decays~\cite{ALEPHtau}
to improve the evaluation of the dispersive integral in eq.~(\ref{eq:hlo}).
Indeed, assuming isospin invariance to hold, the isovector part of the cross
section for $e^+e^-\rightarrow$ hadrons can be calculated via the Conserved
Vector Current ({\small CVC}) relations from $\tau$-decay spectra.  An
updated analysis is presented in ~\cite{DEHZ03}, where $\tau$ spectral
functions are obtained from the results of {\small ALEPH}~\cite{ALEPH02},
{\small CLEO}~\cite{CLEO} and {\small OPAL}~\cite{OPAL}, and
isospin-breaking corrections are applied~\cite{MS88,CEN01,CEN02}. In this
$\tau$-based evaluation, the $2\pi$ and the two $4\pi$ channels are taken
from $\tau$ data up to 1.6 GeV and complemented by $e^+e^-$ data above (the
{\small QCD} prediction for $R(s)$ is employed above 5 GeV).  Note that
$\tau$ decay experiments measure decay rates which are inclusive with
respect to radiative photons.  Their result is
\bea
      \mbox{\cite{DEHZ03}} \qquad\qquad&
      \amu^{\mbox{$\scriptscriptstyle{HLO}$}} \,\,= &  
      7110 \, (50)_{exp} (8)_{rad} (28)_{SU(2)} \times 10^{-11}, 
\label{eq:DEHZ03tau}
\eea
where the quoted uncertainties are experimental, missing radiative
corrections to some $e^+e^-$ data, and isospin violation. This value must be
compared with their $e^+e^-$-based determination in eq.~(\ref{eq:DEHZ04}).
Also the analysis of~\cite{dTY04} includes information from $\tau$
decay. They obtain
\bea
      \mbox{\cite{dTY04}} \qquad\qquad&
      \amu^{\mbox{$\scriptscriptstyle{HLO}$}} \,\,= &  
      7027 \, (47)_{exp} (10)_{rad} \times 10^{-11},~~~~~~~~~
\label{eq:dTY04tau}
\eea
to be compared with their determination in eq.~(\ref{eq:dTY04}).

\begin{figure}[h]
\begin{center}
\includegraphics[width=12cm]{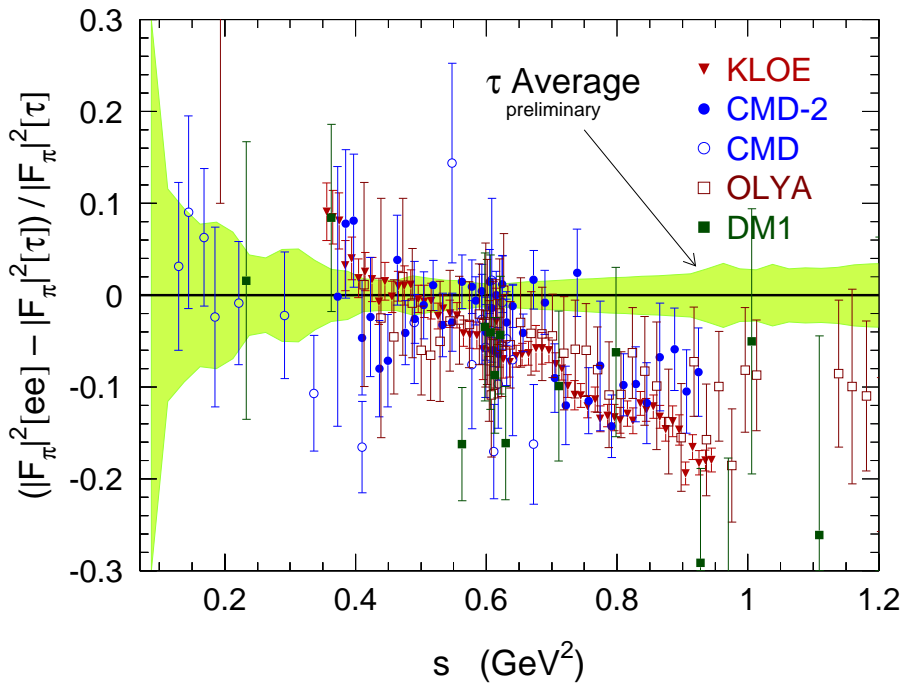}
\caption{{\sf Relative comparison of the $\pi^+\pi^-$ spectral functions
from $e^+e^-$ and isospin-breaking-corrected $\tau$ data, expressed as a
ratio to the $\tau$ spectral functions. The band shows the uncertainty of
the latter. This figure is from ref.~\cite{DEHZ04}.}}
\label{fig:dehz04}
\end{center}
\end{figure}

Although the latest {\small CMD-2} $e^+e^-\rightarrow \pi^+\pi^-$ data are
consistent with $\tau$ data for the energy region below 850 MeV, there is an
unexplained discrepancy for larger energies. This is clearly visible in
fig.~\ref{fig:dehz04}, from ref.~\cite{DEHZ04}, where the relative
comparison of the $\pi^+\pi^-$ spectral functions from $e^+e^-$ and
isospin-breaking-corrected $\tau$ data is illustrated. The same figure also
shows the $\pi^+\pi^-$ spectral functions derived from {\small KLOE}'s 2004
$e^+e^-$ analysis. They are in fair agreement with those of {\small CMD-2}
and confirm the discrepancy with the $\tau$ data.

Among the possible causes of this discrepancy, one may wonder about
inconsistencies in the $e^+e^-$ data, in the $\tau$ data, or in the
isospin-breaking corrections applied to the $\tau$ spectral functions. Given
the good consistency of the {\small ALEPH} and {\small CLEO} data sets, and
the confirmation by {\small KLOE} of the trend exhibited by other $e^+ e^-$
data, further careful investigations of the isospin-violating effects are
clearly warranted -- see the interesting studies in~\cite{dTY04, DEHZ04,
Le02, GJ03, Da03, Mo04}, in particular the discussion of the possible
difference between the masses and the widths of neutral and charged
$\rho$-mesons. Until we reach a better understanding of this problem, it is
probably safer to discard information from $\tau$ decays for the evaluation
of $\amu^{\mbox{$\scriptscriptstyle{HLO}$}}$~\cite{DEHZ04}.

\subsection{Higher-order Hadronic Contributions}
\label{subsec:HHO}

\begin{figure}[h]
\begin{center}
\includegraphics[width=14cm]{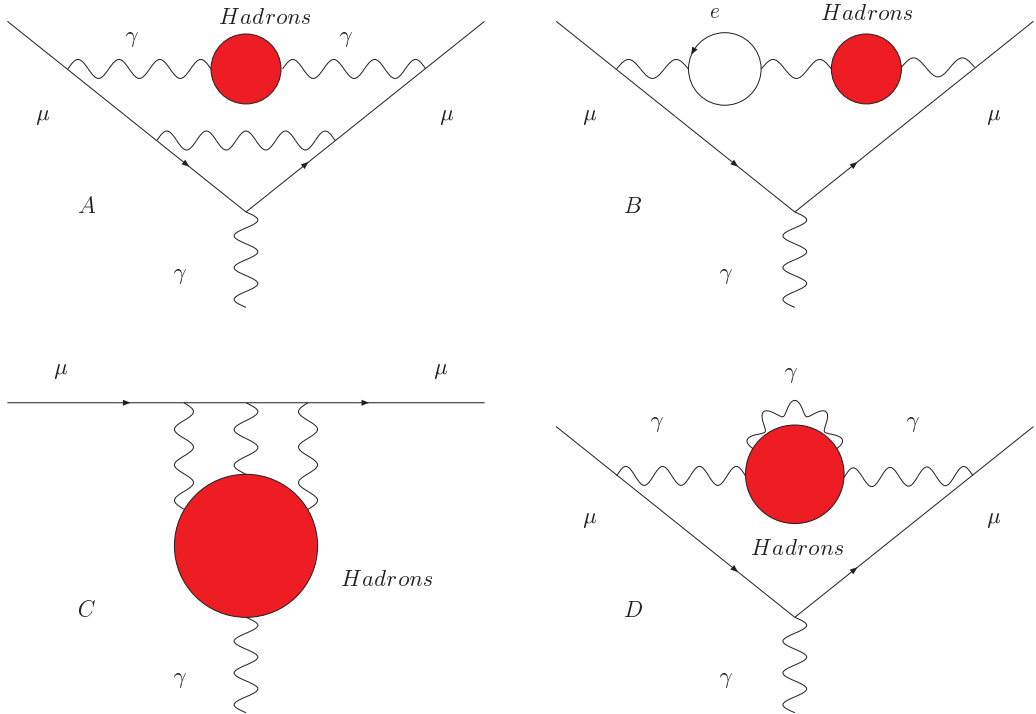}
\caption{{\sf Some of the higher-order hadronic diagrams contributing to
$\amu$.}}
\label{fig:hho}
\end{center}
\end{figure}

We will now briefly discuss the $O(\alpha^3)$ hadronic contribution to
the muon $g$$-$$2$, $\amu^{\mbox{$\scriptscriptstyle{HHO}$}}$, which can be
divided into two parts:
\be
     \amu^{\mbox{$\scriptscriptstyle{HHO}$}}=
     \amu^{\mbox{$\scriptscriptstyle{HHO}$}}(\mbox{vp})+
     \amu^{\mbox{$\scriptscriptstyle{HHO}$}}(\mbox{lbl}).
\ee
The first term is the $O(\alpha^3)$ contribution of diagrams containing
hadronic vacuum polarization insertions, including, among others, those
depicted in figs.~\ref{fig:hho} $A$ and $B$. The second one is the
light-by-light contribution, shown in fig.~\ref{fig:hho} $C$. Note that the
$O(\alpha^3)$ diagram in fig.~\ref{fig:hho} $D$ has already been included in
the leading-order hadronic contribution
$\amu^{\mbox{$\scriptscriptstyle{HLO}$}}$ although, unsatisfactorily, using
{\em scalar} {\small QED} (see discussion in sec.~\ref{subsec:HLO}). In
recent years, $\amu^{\mbox{$\scriptscriptstyle{HHO}$}}(\mbox{vp})$ was
evaluated by Krause~\cite{Kr96} and slightly updated in~\cite{ADH98}. Its
latest value is~\cite{HMNT03}
\be
      \amu^{\mbox{$\scriptscriptstyle{HHO}$}}(\mbox{vp})=
      -97.9 \, (0.9)_{exp} (0.3)_{rad} \times 10^{-11}.
\label{eq:hhovp}
\ee
This result was obtained using the same hadronic $e^+ e^-$ annihilation data
described in sec.~\ref{subsec:HLO}. It changes by about $-3\times 10^{-11}$
if hadronic $\tau$-decay data (again, see sec.~\ref{subsec:HLO}) are used
instead~\cite{DM04}.

%

The hadronic light-by-light contribution changed sign already three times in
its troubled life. Contrary to
$\amu^{\mbox{$\scriptscriptstyle{HHO}$}}(\mbox{vp})$, it cannot be expressed
in terms of experimental observables determined from data and its evaluation
therefore relies on purely theoretical considerations. The estimate of the
authors of~\cite{Ny03, KN01,KNPdR01}, who uncovered in 2001 a sign error in
earlier evaluations, is
\bea
      \mbox{\cite{Ny03}} \qquad\qquad&
      \amu^{\mbox{$\scriptscriptstyle{HHO}$}}(\mbox{lbl})\,\,= &  
      +\,80\,(40)\times 10^{-11}.
\label{eq:hholblNy}
\eea
Earlier determinations now agree with this result~\cite{HK01,BPP01}. Further
studies include~\cite{BCM01,RW02}. At the end of 2003 a higher value was
reported in~\cite{MV03},
\bea
      \mbox{\cite{MV03}} \qquad\qquad&
      \amu^{\mbox{$\scriptscriptstyle{HHO}$}}(\mbox{lbl})\,\,= &  
      +\,136\,(25)\times 10^{-11}.
\label{eq:hholblMV}
\eea
It was obtained by including short-distance {\small QCD} constraints
previously overlooked.  Further independent calculations would provide an
important check of this result for
$\amu^{\mbox{$\scriptscriptstyle{HHO}$}}(\mbox{lbl})$, a contribution whose
uncertainty may become the ultimate limitation of the {\small SM} prediction
of the muon $g$$-$$2$.

\section{The Standard Model Prediction vs.\ Measurement}
\label{sec:COMP}

We now have all the ingredients to derive the {\small SM} prediction for
$\amu$:
\be
    \amu^{\mysmall SM} = 
         \amu^{\mysmall QED} +
         \amu^{\mysmall EW}  +
	 \amu^{\mbox{$\scriptscriptstyle{HLO}$}}  +
	 \amu^{\mbox{$\scriptscriptstyle{HHO}$}}(\mbox{vp})  +
	 \amu^{\mbox{$\scriptscriptstyle{HHO}$}}(\mbox{lbl}).
\label{eq:sm}
\ee
For convenience, I collect here the values of each term from
eqs.~(\ref{eq:qed}, \ref{eq:ew}, \ref{eq:DEHZ04}--\ref{eq:dTY04tau},
\ref{eq:hhovp}--\ref{eq:hholblMV}):

\renewcommand{\arraystretch}{1.3}
$$
\begin{array}{llclr}
      \mbox{[this article]} \qquad\qquad &
      \amu^{\mbox{$\scriptscriptstyle{QED}$}}   &= & 
      116 \, 584 \, 718.8 \,(0.5)  &\times 10^{-11} \\  
      \mbox{\cite{CMV03}} \qquad\qquad &
      \amu^{\mbox{$\scriptscriptstyle{EW}$}}    &= &
      154(1)(2)                       &\times 10^{-11}        \\
      \mbox{\cite{DEHZ04}} \qquad (e^+e^-) &
      \amu^{\mbox{$\scriptscriptstyle{HLO}$}} &= & 
      6934 \, (53)_{exp} (35)_{rad} &\times 10^{-11}             \\
      \mbox{\cite{J03}}    \qquad (e^+e^-) &
      \amu^{\mbox{$\scriptscriptstyle{HLO}$}} &= & 
      6948 \, (86) &\times 10^{-11}                              \\
      \mbox{\cite{ELZ04}}  \qquad (e^+e^-) &   
       \amu^{\mbox{$\scriptscriptstyle{HLO}$}}&= & 
      6934 \, (92) &\times 10^{-11} \\
      \mbox{\cite{HMNT03}} \qquad (e^+e^-) &
      \amu^{\mbox{$\scriptscriptstyle{HLO}$}} &= & 
      6924 \, (59)_{exp} (24)_{rad} &\times 10^{-11}             \\
      \mbox{\cite{dTY04}}  \qquad (e^+e^-) &
      \amu^{\mbox{$\scriptscriptstyle{HLO}$}} &= & 
      6944 \, (48)_{exp} (10)_{rad} &\times 10^{-11}            \\  
      \mbox{\cite{DEHZ03}} \qquad (\tau) &
      \amu^{\mbox{$\scriptscriptstyle{HLO}$}} &= &  
      7110 \, (50)_{exp} (8)_{rad} (28)_{SU(2)} &\times 10^{-11} \\  
      \mbox{\cite{dTY04}} \qquad  (e^+e^-, \tau)  &
      \amu^{\mbox{$\scriptscriptstyle{HLO}$}} &= &  
      7027 \, (47)_{exp} (10)_{rad}     &\times 10^{-11}        \\ 
      \mbox{\cite{HMNT03}} \qquad (e^+e^-) &
      \amu^{\mbox{$\scriptscriptstyle{HHO}$}}(\mbox{vp})\,\,&= & 
      -97.9 \, (0.9)_{exp} (0.3)_{rad} &\times 10^{-11}          \\
      \mbox{\cite{DM04}} \qquad (\tau) &
      \amu^{\mbox{$\scriptscriptstyle{HHO}$}}(\mbox{vp})\,\,&= & 
      -101 \, (1) &\times 10^{-11}          \\
      \mbox{\cite{Ny03}} \qquad\qquad   &
      \amu^{\mbox{$\scriptscriptstyle{HHO}$}}(\mbox{lbl})\,\,&= &  
      80\,(40)                         &\times 10^{-11}          \\
      \mbox{\cite{MV03}} \qquad\qquad   &
      \amu^{\mbox{$\scriptscriptstyle{HHO}$}}(\mbox{lbl})\,\,&= &  
      136\,(25)                        & \times 10^{-11}            
\end{array}
$$
\renewcommand{\arraystretch}{1}

\noindent
The values I obtain for $\amu^{\mysmall SM}$ are shown in the first column
of table~\ref{tab:EXPvsSM}. The values employed for
$\amu^{\mbox{$\scriptscriptstyle{HLO}$}}$ are indicated by the reference in
the last column. I used the latest value available for the hadronic
light-by-light contribution
$\amu^{\mbox{$\scriptscriptstyle{HHO}$}}(\mbox{lbl})= 136\,(25)\times
10^{-11}$~\cite{MV03}. Errors were added in quadrature.

The latest measurement of the anomalous magnetic moment of negative muons by
the experiment {\small E821} at Brookhaven is~\cite{BNL04}
\be 
      a_{\mu^-}^{\mbox{$\scriptscriptstyle{EXP}$}} = 
	  116 \, 592 \, 140 \, (80)(30) \times 10^{-11},
\label{eq:bnl04}
\ee
where the first uncertainty is statistical and the second is systematic.
This result is in good agreement with the average of the measurements of the
anomalous magnetic moment of positive muons~\cite{BNL00,BNL01,BNL02,oldEXP},
as predicted by the {\small CPT} theorem~\cite{Hu03}. The present world
average experimental value is~\cite{BNL04}
\be
    \amu^{\mbox{$\scriptscriptstyle{EXP}$}}  = 
               116 \, 592 \, 080 \, (60) \times 10^{-11} 
	       \quad (0.5~\mbox{ppm}).
\label{eq:exp}
\ee

The comparison of the {\small SM} results with the present experimental
average in eq.~(\ref{eq:exp}) gives the discrepancies
$(\amu^{\mbox{$\scriptscriptstyle{EXP}$}}-
\amu^{\mbox{$\scriptscriptstyle{SM}$}})$ listed in the second column of
table~\ref{tab:EXPvsSM}. The number of standard deviations, shown in the
third column, spans a wide range from 0.7 to 2.8. Somewhat higher
discrepancies, shown in parentheses in the third column, are obtained if the
hadronic light-by-light contribution
$\amu^{\mbox{$\scriptscriptstyle{HHO}$}}(\mbox{lbl})= 80\,(40)\times
10^{-11}$~\cite{Ny03} is used instead of
$\amu^{\mbox{$\scriptscriptstyle{HHO}$}}(\mbox{lbl})= 136\,(25)\times
10^{-11}$~\cite{MV03}, with the number of standard deviations spanning the
range $[1.3-3.2]$ instead of $[0.7-2.8]$. Note that the entries of the first
row in table~\ref{tab:EXPvsSM} are based on the preliminary result for
$\amu^{\mbox{$\scriptscriptstyle{HLO}$}}$ of ref.~\cite{DEHZ04}, which
already includes the recent data from {\small KLOE} and updates the one of
ref.~\cite{DEHZ03}, shifting it down by $29 \times 10^{-11}$. As two thirds
of this shift are due to the inclusion of the {\small KLOE} data, it is
possible that eventually also the $\amu^{\mbox{$\scriptscriptstyle{HLO}$}}$
results of refs.~\cite{J03,ELZ04,HMNT03,dTY04} will undergo some
decrease as a consequence of this inclusion, thus increasing the
corresponding $\amu^{\mbox{$\scriptscriptstyle{EXP}$}}-
\amu^{\mbox{$\scriptscriptstyle{SM}$}}$ discrepancies.
\begin{table}[h] 
\begin{center}
\renewcommand{\arraystretch}{1.1}
\begin{tabular}{|l|l|l|l|}
\hline 
\hline 
$\amu^{\mbox{$\scriptscriptstyle{SM}$}} \times 10^{11}$   & 
$(\amu^{\mbox{$\scriptscriptstyle{EXP}$}}-
  \amu^{\mbox{$\scriptscriptstyle{SM}$}})\times 10^{11}$  & 
$\sigma$                                                  &
{\small HLO} Reference                                    \\
\hline 
116591845 \,(69) & 235 \,(91) & 2.6 ~~~(3.0) & 
      \mbox{\cite{DEHZ04}}~~~$(e^+e^-)$\\ 
116591859 \,(90) & 221 \,(108)& 2.1 ~~~(2.5) & 
      \mbox{\cite{J03}}~~~$(e^+e^-)$\\ 
116591845 \,(95) & 235 \,(113)& 2.1 ~~~(2.5) & 
      \mbox{\cite{ELZ04}}~~~$(e^+e^-)$\\ 
116591835 \,(69) & 245 \,(91) & 2.7 ~~~(3.1) & 
      \mbox{\cite{HMNT03}}~~~$(e^+e^-)$\\ 
116591855 \,(55) & 225 \,(81) & 2.8 ~~~(3.2) & 
      \mbox{\cite{dTY04}}~~~$(e^+e^-)$\\ 
\hline 
116592018 \,(63) &  62 \,(87) & 0.7 ~~~(1.3) & 
      \mbox{\cite{DEHZ03}}~~~$(\tau)$\\ 
116591938 \,(54) & 142 \,(81) & 1.8 ~~~(2.3)&
      \mbox{\cite{dTY04}}~~~$(e^+e^-,\tau)$\\
\hline 
\hline
\end{tabular}
\end{center} 
\caption{{\sf {\small SM} predictions for $\amu$ compared with the current
  measured world average value. See text for details.}}
\label{tab:EXPvsSM}   
\end{table} 
%
%

\section{Conclusions}
\label{sec:CONC}

In the previous sections I presented an update and a review of the
contributions to the {\small SM} prediction for the muon $g$$-$$2$.  What
should we conclude from the wide spectrum of results obtained in
sec.~\ref{sec:COMP}?  The discrepancies in table~\ref{tab:EXPvsSM} between
recent {\small SM} predictions and the current world average experimental
value range from 0.7 to 3.2 standard deviations, according to the values
used for the leading-order and light-by-light hadronic contributions.
In particular, the contribution of the hadronic vacuum polarization depends
on which of the two data sets, $e^+e^-$ collisions or $\tau$ decays, are
employed.

This puzzling discrepancy between the $\pi^+\pi^-$ spectral functions from
$e^+e^-$ and isospin-breaking-corrected $\tau$ data could be caused by
inconsistencies in the $e^+e^-$ data, in the $\tau$ data, or in the
isospin-breaking corrections applied to the latter.  Given the fair
agreement between the {\small CMD-2} and {\small KLOE} $e^+e^-$ data, and
the good consistency of the {\small ALEPH} and {\small CLEO} $\tau$ spectral
functions, it is clear that further careful investigations of the isospin
violations are highly warranted. Indeed, the question remains whether all
possible isospin-breaking effects have been properly taken into
account. Until we reach a better understanding of this problem, it is
probably safer to discard information from hadronic $\tau$
decays~\cite{DEHZ04}. (Of course, discarding $\tau$ data information still
leaves us with the problem of their discrepancy, a troublesome issue on its
own, independent of the calculation of the muon $g$$-$$2$.) If $e^+e^-$
annihilation data are used to evaluate the leading hadronic contribution,
the {\small SM} prediction of the muon $g$$-$$2$ deviates from the present
experimental value by 2--3 standard deviations.

The measurement of the muon $g$$-$$2$ by the {\small E821} experiment at the
Brookhaven Alternating Gradient Synchrotron, with an impressive relative
precision of 0.5 ppm, is still limited by statistical errors rather than
systematic ones.  A new experiment, {\small E969}, has been approved (but
not yet funded) at Brookhaven in September 2004~\cite{E969}. Its goal would
be to reduce the present experimental uncertainty by a factor of 2.5 to
about 0.2 ppm ($\pm 23 \times 10^{-11}$). A letter of intent for an even
more precise $g$$-$$2$ experiment was submitted to {\small J-PARC} with the
proposal to reach a precision below 0.1 ppm~\cite{JPARC}.  While the
theoretical predictions for the {\small QED} and {\small EW} contributions
appear to be ready to rival these precisions, much effort will be needed in
the hadronic sector to test $\amu^{\mysmall SM}$ at an accuracy comparable
to the experimental one. Such an effort is certainly well motivated by the
excellent opportunity the muon $g$$-$$2$ is providing us to unveil or
constrain ``new physics'' effects.

\section*{Acknowledgments}

It is a pleasure to thank G.~Colangelo, J.~Gasser, H.~Leutwyler,
P.~Minkowski, A.~Rusetsky and I.~Scimemi for many useful discussions of the
topic presented here. Several comments and remarks reported in this paper
grew out of our ongoing collaboration. I also wish to thank M.~Davier,
A.~Ferroglia, T.~Kinoshita, A.~Pich, F.~Yndur\'{a}in and O.V.~Zenin for very
helpful comments on a number of points. The exact formulae for the
evaluation of the three-loop {\small QED} coefficients were kindly provided
by S.~Laporta. All diagrams were drawn with {\tt Jaxodraw}~\cite{Jax}.


\end{document}